\documentclass[useAMS,usenatbib, usegraphicx]{mn2e}
\usepackage{epsfig}
\usepackage{graphicx}
\usepackage{amsmath}
\usepackage{amssymb}
\usepackage{natbib}
\newcommand{\infinity}{{\infty}}
\newcommand{\apj}{ApJ}

\newcommand{\mnras}{MNRAS}

\newcommand{\apjs}{ApJS}

\newcommand{\aap}{A\&A}

\begin{document}

\title[Clustering Around Early Galaxies]{Verifying the Identity of High-Redshift Massive Galaxies Through the Clustering of Lower Mass Galaxies Around Them}

\author[Mu{\~n}oz \& Loeb]{Joseph A.\ Mu{\~n}oz \thanks{E-mail: jamunoz@cfa.harvard.edu} and Abraham Loeb \thanks{E-mail: aloeb@cfa.harvard.edu}\\Harvard-Smithsonian Center for Astrophysics, 60 Garden St., MS 10, Cambridge, MA 02138, USA\\}

\maketitle

\begin{abstract}

Massive high-redshift galaxies form in over-dense regions where the probability of forming other galaxies is also strongly enhanced.  Given an observed flux of a galaxy, the inferred mass of its host halo tends to be larger as its inferred redshift increases.  As the mass and redshift of a galaxy halo increase, the expected clustering of other galaxies around it gets stronger.  It is therefore possible to verify the high-redshift identity of a galaxy (prior to an unambiguous spectral identification) from the clustering of other galaxies around it.  We illustrate this method for the massive galaxy suggested by \citet{Mobasher05} to be at redshift $z \! \sim \!6.5$.  If this galaxy were to exist at $z \! \sim \! 6.5$, there should have been a mean of $\sim \! 10$ galaxies larger than a hundredth of its mass and having z-band magnitudes less than $\sim \!25$ detected as i-dropouts in the HUDF.  We calculate an approximate probability distribution for neighbor galaxies and determine that there is less than a $\sim \! 0.3\%$ chance of detecting no massive neighbor galaxies.  The lack of other massive $z \! \sim \! 6.5$ galaxies in the HUDF image argues that the \citet{Mobasher05} galaxy is instead a low redshift interloper.  We generalize our results to other galaxy masses and redshifts.

\end{abstract}


\section{Introduction}

The purported detection of a very massive galaxy, HUDF-JD2, at a redshift of $z \! \sim \! 6.5$ by \citet{Mobasher05} provides a critical test for the standard paradigm of galaxy formation in a $\Lambda$CDM cosmology.  However, the lack of an unambiguous spectral identification allows this galaxy to be a $z \!\sim \! 2$ interloper in which obscuration by dust mimics the Lyman break \citep{Dunlop07,Chary07}.  If the galaxy is at $z \!\sim\! 6.5$, the estimated mass of the halo containing this galaxy is $M_{JD2}\!\sim\!2\times10^{13}\,M_{\odot}$ for a reasonable star formation efficiency, $f_{\star}$, of $10\%$.  In the concordance cosmological model, the expected number of such galaxies in the Hubble Ultra Deep Field (HUDF) field-of-view is less than $10^{-6}$ \citep{BL06}.  Since the expected probability of finding such a galaxy in the HUDF is extremely low, its redshift identification is of great importance for testing the standard cosmological model.

Here we point out that, despite the low average abundance of galaxies like
JD2 (due to its large mass and high redshift), such a galaxy cannot exist
alone.  The large-scale over-density implied by its existence naturally
results in neighboring halos over and above what would be expected from
random fluctuations in the average galaxy population.  The surrounding
over-dense region behaves as if it is part of a closed universe, in which
the formation of all galaxies occurs earlier.  We approach this problem
analytically in this {\it Paper}.  In addition to the deeper fundamental
understanding gained from such a treatment, the extreme rarity of objects
like JD2 make a statistical analysis using numerical simulations difficult.
In \S \ref{Met}, we show how the excursion set formalism can be used to
calculate an approximate probability distribution for the number of
neighbors around massive galaxies.  We then calculate, in particular, the
expected clustering of bright galaxies around the \citet{Mobasher05} galaxy
in \S\ref{JD2}, and explore the dependence of our results on the star
formation efficiency, duty cycle, and power-spectrum normalization in
\S\ref{param}.  In \S\ref{future}, we generalize these results to other
halo masses and redshifts.  Finally, \S\ref{discussion} summarizes our main
conclusions.

Unless otherwise noted, we assume a flat, $\Lambda$CDM model for the universe with the {\it WMAP3} cosmological parameters \citep{Spergel07}.

\section{Method}\label{Met}

We assume the simple model for Lyman-break Galaxies (LBGs) considered by
\citet{SLE07}, which associates LBGs with merger-activated star formation
in dark-matter halos and includes suppression of the star formation
efficiency in low-mass halos by supernova feedback.  In the model, the star
formation duty cycle, $\epsilon_{DC}$, gives the fraction of halos that
contain active star formation.  This fraction has recently been calibrated
by the measured luminosity function of LBGs at $z\!\sim\!6$ to a best-fit value
with $1\!-\!\sigma$ errors of $\epsilon_{DC}=0.25_{-0.09}^{+0.38}$
\citep{SLE07}.  Here we adopt a conservative value of $\epsilon_{DC}=0.14$,
in accordance with our assumed star-formation efficiency of $f_\star=10\%$
(which matches the fraction of $\Omega_b$ in stars today).  The remaining
fraction, $1-\epsilon_{DC}$, of halos at $z\! \sim \!6$ will not be detected as
i-dropout LBGs and may include a population of post-starburst galaxies
similar to JD2 itself.  For our calculations, we assume no variation of the
duty cycle over the redshift range of the selection function of the HUDF.

Halos with active star formation do not constitute a fair sample of the total halo population.  \citet{ST03} show that these halos have undergone substantial accretion in their recent past giving them an extra ``temporal" bias.  While the numerical simulations by \citeauthor{ST03} were done at $z=3$, there is, as yet, no analytical method to predict this extra bias at higher redshift.  Thus, our calculations for the number of neighboring LBGs around massive galaxies at high redshift are lower limits that could be modified in the future with a better understanding of the evolution of the ``temporal" bias with mass and redshift.

According to the excursion set prescription \citep{Zentner07}, if the
linear density fluctuations in the universe are extrapolated to their
values today and smoothed on a comoving scale $R$, a point whose
over-density exceeds a critical value of $\delta_{c}(z) \approx
1.686\,D(z=0)/D(z)$, where $D(z)$ is the linear growth factor at redshift
$z$, belongs to a collapsed object with a mass
$M=(4/3)\,\pi\,\rho_{crit}\,R^3$ if $R$ is the largest scale for which the
criterion is met, where $\rho_{crit}$ is the critical density of the
universe today.  The critical value of the over-density, extrapolated to
today from $z=6.5$, is $\delta_c(z=6.5)\! \sim \!9.6$.  For a Gaussian
random field of initial density perturbations, as indicated by {\it WMAP3}
measurements of cosmic microwave background \citep{Spergel07}, the
probability distribution of the extrapolated and smoothed over-density,
$\delta_R$, is also a Gaussian:
\begin{equation}\label{Q0}
Q_0(\delta_R,S(R))\,d\delta=
\frac{1}{\sqrt{2\,\pi\,S(R)}}\,\exp\left(-\frac{\delta_R^2}{2\,S(R)}\right)\,d\delta,
\end{equation}
with zero mean and a variance given by:
\begin{equation}\label{var}
S\left(R\right)=\int_0^{k_{max}} \frac{dk}{2\,\pi^2}\,k^2\,P\left(k\right),
\end{equation}
where $P\left(k\right)$ is the linear power-spectrum of density fluctuations today as a function of wave-number $k$, and $k_{max}=1/R$.  Since equation (\ref{var}) is a monotonically decreasing function of $R$ (or $M$), the smoothing scale can be uniquely specified by the variance of the over-density field smoothed on that scale.  The critical threshold for collapse introduces a small correction to the probability distribution such that the distribution of $\delta_R$ becomes:
\begin{equation}\label{Q}
Q(\delta_R,S(R))=Q_0(\delta_R,S(R))-Q_0(2\,\delta_c-\delta_R,S(R)).
\end{equation}
The conditional probability distribution of $\delta_1$ on a scale specified by $S_1$ given a value of $\delta_2$ on a scale larger scale specified by $S_2 < S_1$ is:
\begin{equation}\label{Q12}
Q(\delta_1,S_1|\delta_2,S_2)=Q(\delta_1-\delta_2,S_1-S_2).
\end{equation}
Using Bayes Theroem, the conditional probability of $\delta_2$ on a scale $S_2$ given $\delta_1$ on a smaller scale specified by $S_1 > S_2$ is:
\begin{equation}\label{Q21}
Q(\delta_2,S_2|\delta_1,S_1)\,d\delta_1 \propto 
Q(\delta_1,S_1|\delta_2,S_2)\,Q(\delta_2,S_2)\,d\delta_2
\end{equation}
with a constant of proportionality such that the integral of equation (\ref{Q21}) is unity.  Setting $\delta_1=\delta_c(z)$ and $S_1=S(M_1)$, where $M_1$ is the host halo mass of a detected massive galaxy, gives the probability distribution of $\delta_2=\delta_R$ on any comoving scale $R > R_1$, due to the presence of that galaxy, where $R_1$ is the radius corresponding to $M_1$.

However, the excursion set formalism calculates the collapse of objects (a nonlinear effect) by considering the behavior of the linear over-density field extrapolated to the present day.  This method functions entirely in Lagrangian coordinates (which move with the flow) and does not take into account how the over-density field changes in the quasi-linear regime.  Obviously, when a region collapses, matter is pulled in from the surrounding region to fill the void.  Thus, if we assume the existence of JD2 at $z=6.5$, the material in the rest of the HUDF at that redshift would have started outside the region earlier in the universe's history.  

We denote the Lagrangian radius of a region as $R_L$.  Early in the history of the universe, before the region begins to collapse, this radius is equal to the radius of the region in Eulerian coordinates (which do not move with the flow).  As the region collapses, the Eulerian radius shrinks, while $R_L$ remains unchanged.  We denote the final Eulerian radius of the region at the redshift at which it is observed as $R_E$.

The extent of the collapse depends on the magnitude of the over-density in the presence of the massive galaxy whose probability distribution we have just calculated in Lagrangian coordinates.  The more over-dense the region, the larger it would have to be initially to collapse to the same value of $R_E$.  Similarly, a lower value of the over-density would mean that the material inside $R_E$ came from a relatively smaller Lagrangian size.  We would like a mapping, then, between the comoving Eulerian size of a viewed region (such as the HUDF), and the comoving Lagrangian size of the region from where the same material originated in the early universe.  This can be obtained via the spherical collapse model \citep{MW96}.  A spherically symmetric perturbation of Lagrangian radius $R_L$ and over-density $\delta_L > 0$ collapses to a sphere of comoving Eulerian size $R_E$ at redshift $z$ given by:
\begin{equation}\label{RE}
R_E = \frac{3}{10}\frac{1 - \cos\,\theta}{\delta_L}\,\frac{D(z=0)}{D(z)}\,R_L
\end{equation}
\begin{equation}\label{REz}
\frac{1}{1+z} = \frac{3\times6^{2/3}}{20}\frac{\left(\theta -\sin\,\theta\right)^{2/3}}{\delta_L}
\end{equation}
For a fixed value of $R_E$, there is a one-to-one relationship between $R_L$ and the value of $\delta_L$ that collapses $R_L$ to $R_E$.

In the presence of a massive galaxy, the probability distribution of $\delta$ (in the Lagrangian sphere that collapsed to $R_E$) can be computed by considering the possible histories of the region $R_E$ having collapsed from different possible $R_L$'s weighted by the probability of the corresponding value of $\delta_L$ in each $R_L$.  These weights are given by equation (\ref{Q21}).  The resulting probability distribution of $\delta$ (dropping the subscript L) ``seen" in a fixed $R_E$ can be expressed generally as:
\begin{equation}\label{Pdel}
\frac{dP(\delta | M_1)}{d\delta} \propto Q(\delta,R_L(\delta, R_E) | \delta_c(z), R(M_1)),
\end{equation}
where again the constant of proportionality is set so that $\int (dP(\delta | M_1)/d\delta)\,d\delta = 1$.

\begin{figure}
\begin{center}
\includegraphics[width=\columnwidth]{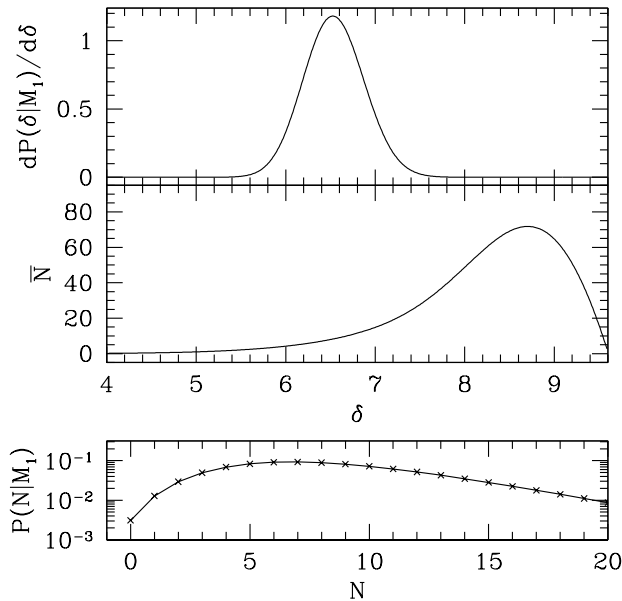}
\caption{\label{Pdist} The upper panel shows the probability distribution
of the over-density $\delta$ (extrapolated to $z=0$) due to cosmic
variance (Eq. \ref{Pdel}) in a spherical region that collapsed to the size
of the HUDF assuming the existence of a $2\times10^{13}\,M_{\odot}$ halo
containing JD2 at $z=6.5$.  The central panel shows, for each value of the
over-density, the resulting number of LBGs in halos with mass above
$2\times10^{11}\,M_{\odot}$ expected in this region.  The final probability
distribution of the number of LBGs in the region above
$2\times10^{11}\,M_{\odot}$ is plotted in the bottom panel, taking into
account both cosmic variance and Poisson fluctuations.  }
\end{center}
\end{figure}

Now that we know the distribution of over-densities in which the massive galaxy sits, we can easily calculate the expected number of neighbor galaxies seen in $R_E$ in the presence of each value for the over-density, $\bar{N}(\delta | M_1)$, and thus, the probability distribution of $\bar{N}$.  \citet{BL04} calculate the mass function of objects in a region of fixed over-density by combining the Sheth-Tormen and Press-Schechter prescriptions in the regimes for which each best fits results from numerical simulations.  Given the relationship between $\delta$ and $R_L$ for a given $R_E$, the over-density  determines the Lagrangian scale size.  We find
\begin{equation}\label{Nvsdel}
\bar{N}(\delta | M_1) = (V(\delta)-V(M_1))\,\epsilon_{DC}\,\int_{m_{min}}^{m_{max}}\frac{dn_{bias}}{dm}(m,z,\delta)\,dm
\end{equation}
\begin{equation}\label{nbias}
\frac{dn_{bias}}{dm}(m,z,\delta) = \frac{dn_{ST}}{dm}(m,z)\frac{f_{PS}(\delta_c(z)-\delta, S(m)-S(\delta))}{f_{PS}(\delta_c(z), S(m))},
\end{equation}
where $dn_{ST}/dm$ is the Sheth-Tormen mass function, $V(\delta)$ is the volume enclosed by the Lagrangian radius specified by $\delta$, and 
\begin{equation}\label{fPS}
f_{PS}(\delta_c(z),S)\,dS =
\frac{\delta_c(z)}{S^{3/2}}\,Q_0(\delta_c(z),S)\,dS
\end{equation}
is the mass fraction at $z$ contained in halos with mass in the range corresponding to $(S,S+dS)$.  The mass limit $m_{max}$ is the mass enclosed by a sphere of today's critical density with radius $R_L$ corresponding to $\delta$.  We ignore the probability that $R_L$, while containing an over-density $\delta<\delta_c(z)$, might be part of a larger collapsed region with $\delta>\delta_c(z)$.  This is a good assumption given the unlikely occurrence of $\delta=\delta_c(z)$ on the scale of $M_1$ in the first place.

The resulting probability distribution of $\bar{N}$, due to cosmic variance, is given by:
\begin{equation}\label{PNbar}
\frac{P(\bar{N} | M_1)}{d\bar{N}} = \frac{dP(\delta(\bar{N}) | M_1)}{d\delta}\,\frac{d\delta}{d\bar{N}},
\end{equation}
where $\delta(\bar{N})$ is the inverse of equation~(\ref{Nvsdel}) and $d\delta/d\bar{N}$ is its derivative.  Poisson fluctuations contribute additional variation in the actual number, $N$, of neighbor galaxies.  The probability, $P(N | M_1)$, of each discrete value of $N$ in the presence of a mass $M_1$ galaxy at high redshift is obtained by convolving a discretized version of equation~(\ref{PNbar}) with the Poisson distribution in the following way:
\begin{equation}\label{PN}
P(N | M_1) = \sum_{\bar{N}=0}^{\infinity} \tilde{P}(\bar{N} | M_1)\,P_{Poisson}(N,\bar{N}),
\end{equation}
where 
\begin{equation}
P_{Poisson}(k,\lambda)=\frac{\lambda^k\,e^{-\lambda}}{k!}, \nonumber
\end{equation}
\begin{equation}
\tilde{P}(\bar{N}=0 | M_1) = \int_0^{0.5} \frac{P(\bar{N} | M_1)}{d\bar{N}}\,d\bar{N}, \nonumber 
\end{equation} 
and for $\bar{N}>0$,
\begin{equation}\label{tilPNbar}
\tilde{P}(\bar{N}| M_1) = \int_{\bar{N}-0.5}^{\bar{N}+0.5} \frac{P(\bar{N}'
| M_1)}{d\bar{N}'}\,d\bar{N}'.
\end{equation}
Equation~(\ref{PN}) can be compared to galaxy counts in surveys.  In particular, if no galaxies in halos above some minimum mass, $m_{min}$, are seen as neighbors to another galaxy in a halo of mass $M_1$, then the quantity $1-P(N=0|M_1)$ is approximately the confidence by which we can rule out either the existence of a halo of mass $M_1$ or the cosmological model.

\section{HUDF-JD2}\label{JD2}

We show results for neighbor i-dropouts around JD2 in the HUDF.  If JD2 is
indeed a massive galaxy at $z\! \sim \! 6.5$, then there should be many
massive galaxies visible around it.  Figure~\ref{Pdist} shows the
distribution of $\delta$ given by equation~(\ref{Pdel}) inside the
Lagrangian patch that collapses to an angular Eulerian size of
$\sim\!115''$, enclosing an area on the sky roughly equivalent to the HUDF
field-of-view.  This angular scale corresponds to a comoving size of $\sim
\!4.7\,{\rm Mpc}$ at $z=6.5$, which assuming an extrapolated over-density
of $\delta=6.5$, encloses within a spherical radius a mass of $\sim
\!10^{14}\,M_{\odot}$.  Also shown are the expected number of neighbors
inside that radius, given each value of the over-density, and the resulting
probability distribution of the number of neighbors that includes both
cosmic variance and Poisson fluctuations.

While the HUDF should be sensitive to LBGs in hosts as small as $\sim\!2\times10^{10}\,M_{\odot}$, the number of random galaxies expected in such hosts without correlations from JD2 is comparable to the number of excess neighbors that result from these correlations.  This creates some ambiguity in detecting the excess over the background.  On the other hand, while the mean number of uncorrelated LBGs in halos as large as $\sim\!2\times10^{12}\,M_{\odot}$ is orders of magnitude smaller than the excess due to JD2, the probability of detecting no such neighbor is not small enough for a lack of a detection to be meaningful.

\begin{figure}
\begin{center}
\includegraphics[width=\columnwidth]{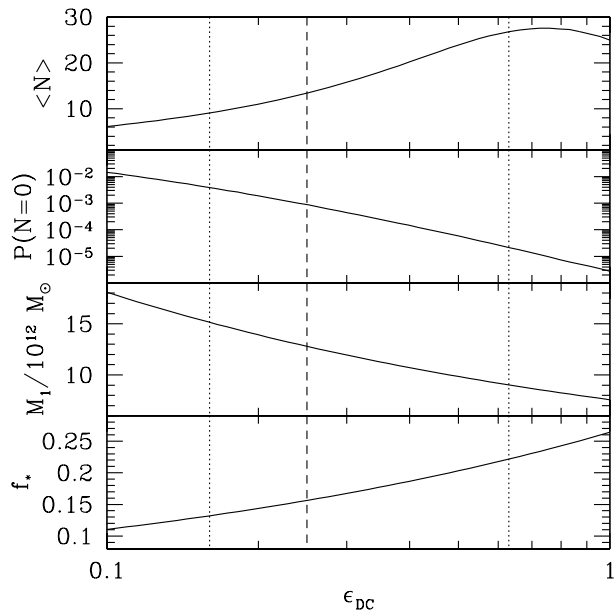}
\caption{\label{neighE} The dependence of the number of neighbor galaxies
on the star formation efficiency and the duty cycle assuming a fixed value
of $2\times10^{12}\,M_{\odot}$ for the stellar mass of JD2.  The upper two
panels show the mean number of neighbor LBGs at $z=6.5$ with a host halo
mass above $0.01 \times M_{JD2}=2\times10^{11}\,M_{\odot}$ in the HUDF due
to the presence of JD2 and the probability of detecting none of these
objects as a function of the duty cycle.  The assumed one-to-one
relationship between the star formation efficiency and the duty cycle given
in equation~(\ref{parameters}) is plotted in the third panel, while the
lower panel shows how the host halo mass of JD2 depends on the duty cycle
through its relationship to the star formation efficiency.  The vertical
dashed line denotes the best-fit value of the duty cycle given by
\citet{SLE07}, while the dotted lines indicate the $1-\sigma$ bounds.
Since the validity of equation~(\ref{parameters}) cannot be verified beyond
the range of the $2-\sigma$ contour, we truncate the plot near its lower
boundary at $\epsilon_{DC}$ of $\sim 0.1$.}

\end{center}
\end{figure}

\begin{figure}
\begin{center}
\includegraphics[width=\columnwidth]{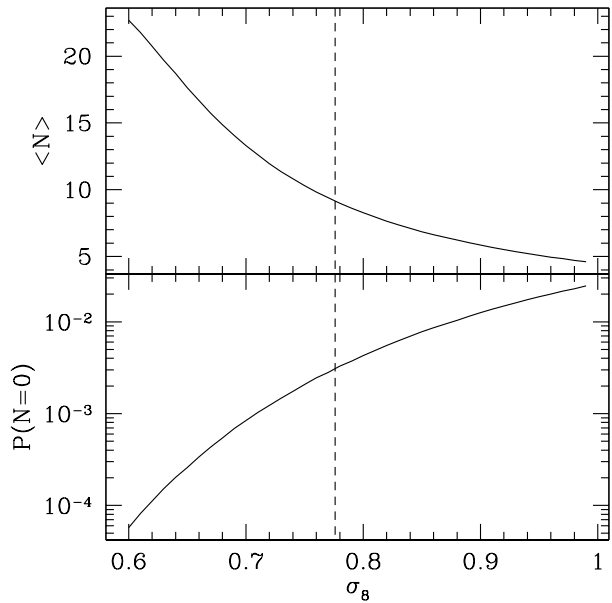}
\caption{\label{neigh8} The dependence of the number of neighbor galaxies on the normalization of the matter power spectrum $\sigma_8$.  The upper panel shows the mean number of neighbor LBGs at $z=6.5$ with a host halo mass above $0.01 \times M_{JD2}$ in the HUDF due to the presence of JD2, while the lower panel indicates the probability of detecting no such objects.  The vertical dashed line denotes the WMAP value used in the rest of the \it{Paper}.}
\end{center}
\end{figure}

Thus, we consider neighbor LBGs in halos with masses above $2\times10^{11}\,M_{\odot}$.  For a duty-cycle $\epsilon_{DC}=0.14$ and star formation efficiency $f_\star=0.1$, these galaxies have luminosities at $1500\,{\rm \AA}$ above $\sim\! 2\times10^{29} {\rm ergs\,s^{-1}\,Hz^{-1}}$ or z-band magnitudes at $z\! \sim \!6.5$ less than $\sim\!25$.  The mean abundance of such galaxies in the absence of JD2 is much less than unity, and indeed, no such objects have been detected in the HUDF \citep{Bouwens06,Bouwens07}.  Figure~\ref{Pdist} shows that there should be a mean of $\sim\! 10$ very bright i-dropout LBGs in hosts with masses larger than within an angular Lagrangian radius of $\sim\!175''$ of JD2, where this is a lower limit due to the fact that LBGs should be more clustered than halos \citep{ST03}.  The probability, given by equation~(\ref{PN}), of detecting no such galaxies in this region is $P(N=0 | M_{JD2})\! \sim \!3\times10^{-4}$.  Thus, we can rule out JD2 at redshift $z\! \sim \!6.5$ with 99.7\% confidence.  Integrating the distribution, we find a less than a $5\%$ chance of detecting fewer than $3$ very bright neighbor galaxies.  Either JD2 is at $z\! \sim \!2$, as allowed by spectral fits, or there is a problem with our assumed cosmological model.

\section{Parameter Dependence}\label{param}

So far, we have assumed a star formation efficiency of $f_{\star}=10\%$, a duty cycle of $\epsilon_{DC}=0.14$, and WMAP3 cosmological parameters with $\sigma_8=0.776$.  We now explore how our results depend on these parameters.

\subsection{$\epsilon_{DC}$ and $f_{\star}$}

Given a stellar mass of $2\times10^{12}\,M_{\odot}$ for JD2 \citep{BL06}, changes in the star formation efficiency will affect the assumed host halo mass.  More efficient star formation will, therefore, result in fewer neighbor halos due to the lower corresponding halo mass.  Meanwhile, changes in the duty cycle will affect the fraction of such halos that are seen as i-dropouts.  Yet, these two parameters are not constrained independently.  Figure 4 of \citet{SLE07} shows very narrow likelihood contours in log-parameter-space.  As a rough approximation, then, we assume that the range allowed by the luminosity function fitting spans a power-law relationship between the star formation efficiency and the duty cycle.  Using the best-fit parameter values and the extremes of the $1-\sigma$ contour, we fit the dependence with a least-squares regression (in log-space) and find a relationship given by:
\begin{equation}\label{parameters}
f_{\star} = A\,\epsilon_{DC}^{\beta},
\end{equation}
where $A=0.264$ and $\beta= 0.378$.

\begin{figure}
\begin{center}
\includegraphics[width=\columnwidth]{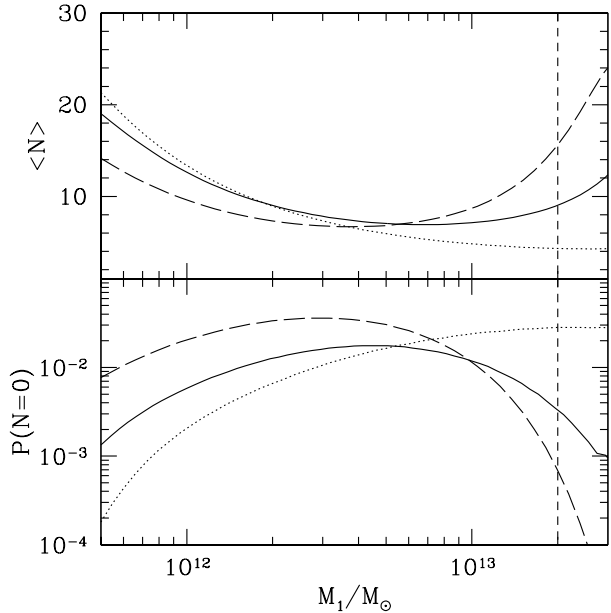}
\caption{\label{neighM} Results for future detections of massive galaxies at fixed $z_1$ as a function of host halo mass $M_{1}$.  The upper panel shows the mean number of neighbor LBGs with a host halo mass above $0.01 \times M_{1}$ within the angular Eulerian distance corresponding to the HUDF, while the probability of finding no such LBGs in the HUDF is plotted in the lower panel.  The dotted, solid, and long-dashed lines denote values for $z_{1} = 5$, $6.5$, and $8$, respectively.  The vertical dashed line denotes the host halo mass of JD2.}
\end{center}
\end{figure}

\begin{figure}
\begin{center}
\includegraphics[width=\columnwidth]{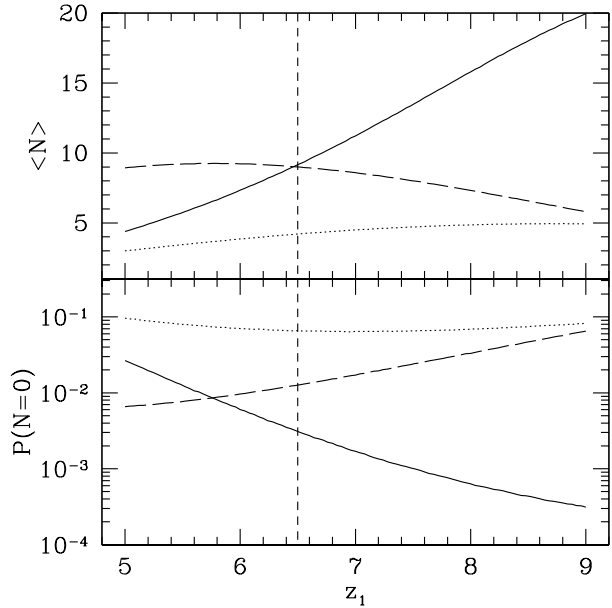}
\caption{\label{neighz} Results for future detections of massive galaxies as a function of redshift $z$.  The upper panel shows the mean number of neighbor LBGs with a host halo mass above $0.01 \times M_{1}$ in the HUDF, while the probability of finding no such LBGs in the HUDF is plotted in the lower panel.  The solid, dotted, and long-dashed lines denote values for $M_{1} = 2\times10^{13}\,M_{\odot}$, $10^{13}\,M_{\odot}$, and $2\times10^{12}\,M_{\odot}$, respectively.  The vertical short-dashed line denotes $z=6.5$, the redshift of JD2.}
\end{center}
\end{figure}

The mean number of neighbor i-dropouts to JD2 in the HUDF with host halo
masses greater than $0.01\,M_1$, denoted $\left<N\right>$, is plotted in
Figure~\ref{neighE} as a function of $\epsilon_{DC}$.  Also shown is the
effect on the probability of detecting no such galaxies, given by setting
$N=0$ in equation (\ref{PN}).  As shown, the effect of varying parameters
on $\left<N\right>$ is only modest.  When considering the range of
$1-\sigma$ errors on $\epsilon_{DC}$, $\left<N\right>$ varies by only a
factor of $\sim\! 2$.  The correlation between the star formation efficiency
and the duty cycle given by equation~(\ref{parameters}), causes these
parameters to moderate each other's effect on the expected number of
neighbors.  While the value of $P(N=0)$ varies more significantly, our
choice of parameters throughout the rest of the paper gives conservative
values for both $\left<N\right>$ and $P(N=0)$.  The best-fit value of
$\epsilon_{DC}=0.25$ derived by Stark et al. (2007) results in an even
smaller probability of detecting no bright neighbor dropouts if JD2 is at
$z=6.5$.

\subsection{$\sigma_8$}

The high over-density in the region surrounding a massive galaxy at high
redshift results from the large density fluctuation required to produce
such a galaxy.  If the over-density must reach $\delta_c(z)$ on the scale
corresponding to the size of the galaxy, then on a smoothing scale only a
little larger, the over-density could not have been very low, since the
contribution from the intervening scales is a Gaussian about zero with a
standard deviation much less than $\delta_c(z)$.  However, varying the
normalization of the matter power-spectrum will change the standard
deviation of this Gaussian contribution.  Reducing the value of $\sigma_8$
will decrease the contribution from these intervening scales and cause the
over-density around massive galaxies to be larger.  This will result in
more neighboring galaxies on average and a lower likelihood of detecting
none.  Conversely, increasing $\sigma_8$ will boost the possible
contribution from intervening scales, shift the probability distribution of
$\delta$ in regions around massive, high-redshift galaxies toward lower
values, and result in lower average number of neighbors and a greater
chance of detecting none.

We explore the dependence of $\left<N\right>$ and $P(N=0)$ on $\sigma_8$ quantitatively in Figure~\ref{neigh8}.  The behavior is just as expected. The dependence is relatively strong given the wide range of proposed values for $\sigma_8$.  However, $P(N=0)$ reaches only $\sim\! 1\%$ at $\sigma_8=0.9$.  

\section{Potential Future Detections}\label{future}

As new surveys continue to probe for ever larger galaxies at ever higher redshifts \citep{McLure06, Rodighiero07, Wiklind07}, it is important to generalize our results to potential future detections of other massive, high-redshift galaxies in the future.  We again assume $(\epsilon_{DC},f_{\star},\sigma_8)=(0.14,0.1,0.776)$ and calculate the expected number of neighbors as for JD2 but now allowing the mass, $M_1$, and redshift, $z_1$, of the halo hosting the detected galaxy to vary.

Figure~\ref{neighM} shows the mean number of neighbor LBGs, $\left<N\right>$, with host halo mass $M_2>0.01\times M_1$ that are within an angular Eulerian separation $\theta=115''$ from a central galaxy with a host halo mass of $M_1$ situated at a fixed $z_1$ as a function of $M_1$.  While the uncorrelated, average abundance of lower mass halos is larger than that of higher mass halos, the plot demonstrates that the average number of neighbors begins to increase as a function of $M_1$, for $z_1=6.5$ and $8$ above $M_1 \! \sim \! 5 \times 10^{12}\,M_{\odot}$ due to the nonlinear increase in halo clustering with mass.  Below this value, the increasing uncorrelated abundance of objects with lower mass causes the number of neighbors to increase with decreasing $M_1$.  At $z_1=5$, $\left<N\right>$ decreases as $M_1$ increases in the entire range considered.  The correlative effects of even a very massive halo are dominated by the uncorrelated behavior at this redshift.  The mean number of neighbor LBGs within $115''$ is greater than unity for all values of $M_1$ and $z_1$ considered.  

The probability of detecting no such neighbors, $P(N=0)$, is also plotted in Figure~\ref{neighM}.  Its dependence on $M_1$ is as expected; $P(N=0)$ increases as $\left<N\right>$ decreases.

In Figure~\ref{neighz}, $\left<N\right>$ and $P(N=0)$ are plotted as functions of $z_1$ at fixed $M_1$ and $\theta=115''$.  While the plots of $\left<N\right>$ and $P(N=0)$ increase and decrease, respectively, with increasing $z_1$ for $M_1=10^{13}\,M_{\odot}$ and $2\times10^{13}\,M_{\odot}$ due to increased clustering with the detected halo, this is not indicated for $M_1=2\times10^{12}\,M_{\odot}$.  As shown in Figure~\ref{neighM}, $d\left<N\right>/dM_1 < 0$ at $M_1=2\times10^{12}\,M_{\odot}$ for all values of $z_1$ considered indicating that clustering is less important for these halos and their neighbors.  This is also seen in the larger values of $\left<N\right>$ for $M_1=2\times10^{12}\,M_{\odot}$ compared to $10^{13}\,M_{\odot}$.  The increase is due to an uncorrelated population of halos with mass $2\times10^{10}\,M_{\odot}<M_2<10^{11}\,M_{\odot}$.  For each set of parameters, a higher value $\left<N\right>$ corresponds to a lower value of $P(N=0)$, as expected.

\section{Discussion}\label{discussion}

We have presented a clear signature for the existence of a massive galaxy at high redshift.  Due to large correlations among massive halos at high redshifts, once such an object is found, it is unlikely to be alone.  On the contrary, the more massive the halo and the higher its redshift, the greater the number of neighbors it has.  This signature is helpful in situations where only photometric data is available or when the spectroscopic redshift identification is ambiguous. For a galaxy of a given observed flux, the inferred halo mass increases as its suggested redshift increases.  It therefore becomes easier to rule out a higher-redshift hypothesis, using dropout techniques to locate the neighboring LBGs that are expected at a similar redshift.

We derived a probability distribution for the number of neighbors around a massive galaxy at high redshift, which led to calculations of the mean number of such neighbors and the probability of detecting no such halos.

In the case of JD2, the number of neighbor galaxies observed in this way can distinguish between the $z\! \sim \!6.5$ interpretation of \citet{Mobasher05} and that of a galaxy at $z\! \sim \!2$ \citep{Dunlop07, Chary07}.  If at $z=6.5$, the mean number of excess i-dropouts in the HUDF with z-band magnitudes less than $\sim\!25$ is predicted to be $\sim\! 10$.  The lack of such bright i-dropouts in the HUDF \citep{Bouwens06, Bouwens07} implies that JD2 cannot be at such high redshift with $99.7\%$ confidence.  Thus, future detections of massive galaxies at high redshift could be verified by looking for bright, massive neighbors.

Additional research into exactly how LBGs form in dark matter halos and how biased they are compared to these halos would firm-up the predictions for the probability distribution of neighbors.  Such work would undoubtedly include various feedback processes and environmental effects on LBG formation.  Previous study indicates that LBGs are more clustered than dark matter halos \citep{ST03}, and our results for the number of neighbors are, thus, lower limits on the true values.

\section{Acknowledgements}

We thank Giovanni Fazio, Kamson Lai, Evan Scannapieco, Dan Stark, and Matt McQuinn, for useful discussions.  JM acknowledges support from a National Science Foundation Graduate Research Fellowship.


\end{document}